\documentclass[runningheads,a4paper]{llncs}

\usepackage{amssymb}
\usepackage{multirow}
\usepackage{graphicx}
\usepackage{array}
\usepackage{subfig}
\usepackage{caption}
\usepackage{url}
\urldef{\mailsa}\path|rshams@uwo.ca|

\begin{document}

\mainmatter
\title{Extracting Information-rich Part of Texts using Text Denoising}
\titlerunning{Extracting Information-rich Part of Texts using Text Denoising}
\author{Rushdi Shams}

\authorrunning{Rushdi Shams}
\institute{Department of Computer Science\\
University of Western Ontario\\London, ON N6A 5B7,Canada\\
\mailsa}

\toctitle{Extracting Information-rich Part of Texts using Text Denoising}
\tocauthor{Rushdi Shams}
\maketitle

\begin{abstract}
\label{abstract}
The aim of this paper is to report on a novel text reduction technique, called \textit{Text Denoising}, that highlights information-rich content when processing a large volume of text data, especially from the biomedical domain. The core feature of the technique, the text readability index, embodies the hypothesis that complex text is more information-rich than the rest. When applied on tasks like biomedical relation bearing text extraction, keyphrase indexing and extracting sentences describing protein interactions, it is evident that the reduced set of text produced by text denoising is more information-rich than the rest.

\end{abstract}


\section{Introduction}
\label{intro}

Often, to test a method's scalability as well as its performance across genres of texts, there is a need to process large volumes of text data in many disciplines of NLP, be it textual relation extraction, summarization or meta-tagging. It has been reported by many researchers \cite{Medelyan:2008}\cite{Witten-et-al:2004} that machine learning as well as rule-based approaches show improvements over their benchmarks with increased training data. However, the use of large volume of data can create several bottlenecks. One is technical---processing large data, like that from biomedical texts, slows down many algorithms; another is even more important---algorithms can exhibit a decreased accuracy because of the noise, which are irrelevant or redundant data for a given classification task, added by information-poor parts of texts.

There are several statistics, like word-level feature \textit{tf--idf} and sentence-level feature \textit{sentence position}, that help identify information richness. Although the degree of use shows their popularity, these features have some serious limitations. For example, \textit{tf--idf} computes document similarity directly in the word-count space which may be slow for large vocabularies and sentence position is useful for summarization but is superficial in relation extraction. In other words, they are either task-specific and/or domain-specific measures. 

Text readability has multivariate features that consider many attributes like length of paragraph, words and sentences, and number of polysyllabic and monosyllabic words. In this paper, I report a text reduction technique called \textit{Text Denoising} that reduces text data based on text readability, especially from the biomedical domain, to that which is more information-rich by removing most of the noise. The reduced text is also expected to be task-independent and informative enough to improve accuracy of NLP tools across disciplines. 

\section{Proposed Method}
\label{proposed} 

Among text readability scores, the following five measures are considered as yardsticks--- Fog Index (hereinafter, FI) \cite{Fog:1969}, Flesch reading ease score (FRES) \cite{Flesch:1948}, Smog Index \cite{Mclaughlin:1969}, Forcast Index \cite{Caylor-et-al:1973}, and Flesch-Kincaid readability index (FKRI) \cite{Kincaid-et-al:1975}. The choice of using text readability as an \textit{information richness statistic} is motivated by the results of an experiment by Duff and Kabance \cite{Duffy:1982}. In their experiment, a passage with no more than two phrases were converted into primer prose and FI was applied to test its readability. They found that the score was low (i.e., the prose was extremely easy to read). The authors concluded that easy texts obscure the relationships and ideas as they de-emphasize both. In contrast, difficult texts emphasize relationships and ideas yielding low readability. I suggest that the describing of biomedical relations, meta information, etc. lengthens sentences as well as increases the use of polysyllabic words which are the two principal components of many of the readability indexes.

Both rule-based and machine learning-based versions of \textit{Text Denoising} are based on this principle that use text readability as a key feature and applies it at the sentence-level to identify those sentences within a text, called denoised text, where content information, such as biomedical relations, is more likely to occur. The rest of the text is called noise text. I am interested to observe the effect of using text denoising on different tasks and genres of text.

\section{Text Denoising on Relation Extraction}
\label{relex}

I developed a corpus of $24$ texts that describe four pairs of related MeSH C and MeSH D concepts reported by Perez \textit{et al.} \cite{Perez-et-al:2006}. I applied the rule-based version of text denoising on these texts to extract related biomedical concepts. The only rule I set for this task was to extract 30\% of the low-readability sentences from the texts according to their FI score. This threshold is termed as the \textit{denoising threshold} and the texts extracted are called \textit{denoised texts}; the rest is called \textit{noise text}. This threshold point was set heurisitically considering the stability in the frequency of appearance of the related concepts in the corpus. Other than 30\%, the results with different denoising thresholds ranging from 10\% to 50\%, however, was not satisfactory. I ranked the pairs of concepts present in the denoised texts using their frequency. Most of the concept pairs with higher ranks, however, did not contain any semantic relations according to UMLS semantic relation network. Therefore, I re-ranked the pairs according to their positive predictive value (PPV) (similar to \textit{precision} measure used in information retrieval evaluation tasks) and sensitivity. The pairs of concepts found from this re-ranking showed a convincing accuracy of 75\% (ratio of semantically related concepts to total) against the output of the UMLS semantic relation network. Table \ref{table1} shows an output from a paper on one of the four pairs of concepts. Of note, I found that the noise texts did not have any related biomedical concepts. The detailled experimental setup and results are reported by Shams and Mercer \cite{Shams:2011}. 

Later, I performed an experiment with four other readability scores mentioned in Section \ref{proposed} on the same corpus. A comparative result showed that FI outperformed the other indexes by extracting more meaningful relations \cite{Shams:WIS2012}. I also analyzed the performance of the indexes considering the performance of FI as a benchmark. Table \ref{micro-average} and \ref{macro-average} show that the SMOG index is a close second to FI followed by FKRI, while FRES and FORCAST performed poorly. It can also be noted that the SMOG index, like FI, uses the core measure of \textit{complex words} which reveals the fact that the measure of complex word fits best for text denoising and biomedical relation extraction. 

\begin{table}[tb!]
\centering
{\footnotesize
\begin{tabular}{>{\centering\arraybackslash}p{1cm}>{\centering\arraybackslash}p{2cm}>{\centering\arraybackslash}p{1.5cm}}
\hline
Rank & Related Concepts & Semantic Relation\\
\hline
1	&\multicolumn{1}{l}{Ischemia-Glutamate}&Yes\\

2	&\multicolumn{1}{l}{Levels-Ischemia}&No\\

3	&\multicolumn{1}{l}{Levels-Glutamate}&Yes\\

4	&\multicolumn{1}{l}{Glutamate-Neurons}&Yes\\

5	&\multicolumn{1}{l}{10min-Ischemia}&Yes\\

6	&\multicolumn{1}{l}{Glutamate-CA4}	&Yes\\

7	&\multicolumn{1}{l}{Increase-Glutamate}		&Yes\\

8	&\multicolumn{1}{l}{10min-Glutamate}		&No\\

9	&\multicolumn{1}{l}{Ischemia-5min}		&Yes\\

9	&\multicolumn{1}{l}{Glutamate-5min}		&No\\
\hline
\end{tabular}

\caption{Extracted related concepts for a paper on Ischemia and Glutamate}
\label{table1}}

\end{table}

\begin{table*}[b]
\centering
\subfloat[]{
\centering
\begin{tabular}{>{\centering\arraybackslash}p{1.5cm}>{\centering\arraybackslash}p{1.2cm}>{\centering\arraybackslash}p{1.2cm}>{\centering\arraybackslash}p{1.2cm}}
\hline
Score  &Precision &Recall &F-Score\tabularnewline
\hline
\multicolumn{1}{l}{SMOG} &95.83  & 82.14  & 88.46 \tabularnewline
\multicolumn{1}{l}{FKRI} &88.89   &82.76  &85.71 \tabularnewline
\multicolumn{1}{l}{FRES} &82.61  &65.52  &73.08 \tabularnewline
\multicolumn{1}{l}{FORCAST} &81.82 &62.07  &70.59 \tabularnewline
\hline
\end{tabular}
\label{micro-average}
}
\subfloat[]{
\centering
\begin{tabular}{>{\centering\arraybackslash}p{1.5cm}>{\centering\arraybackslash}p{1.2cm}>{\centering\arraybackslash}p{1.2cm}>{\centering\arraybackslash}p{1.2cm}}
\hline
Score  &Precision &Recall &F-Score\tabularnewline
\hline
\multicolumn{1}{l}{SMOG} &96.88  &82.60  &89.16 \tabularnewline

\multicolumn{1}{l}{FKRI} &89.73  &82.60  &86.01 \tabularnewline

\multicolumn{1}{l}{FRES} &80.83  &65.63  &72.44 \tabularnewline

\multicolumn{1}{l}{FORCAST} &77.88 &61.61  &68.72 \tabularnewline
\hline
\end{tabular}
\label{macro-average}
}
\label{macro-micro}
\caption[]{\subref{micro-average} Micro-average and \subref{macro-average} macro-average precision, recall and F-Score of the indexes on biomedical relation extraction}
\end{table*}

\section{Text Denoising on Keyphrase Extraction}
\label{keyphrase}

I investigated the usability of denoised texts as training data for machine learning-based keyphrase indexers called KEA \cite{Witten-et-al:2004}, KEA++ \cite{Medelyan:2008} and Maui \cite{Olena:2009}. I applied the indexers with their classifiers induced from denoised training data on three datasets, namely FAO-780, CERN-290 and NLM-500. These datasets are composed of texts from the domains of agriculture, physics and biomedical science. I compared the result with their benchmark performances that were achieved by using the full-text training data. Convincingly, in a 10-fold cross validation experiment, both KEA and KEA++, with their classifiers induced from denoised training data, outperformed their respective benchmark F-scores \cite{Shams:ICADL2012}. Maui, on the other hand, had mixed results and its denoised text induced classifier performs comparably with its benchmark \cite{Shams:2012}. The F-Scores are listed in Table \ref{KEA}, \ref{KEA++} and \ref{Maui} where a \textit{t-value} greater than or equal to $2.26$ indicates the statistical significance of the results at 95\% confidence. Of note, unlike the fixed denoising threshold of 30\% for relation extraction, I found that to get bias-free classifiers for the indexers, the denoising threshold point needed to be varied (usually between 30\%--70\%) for different genres of texts. This outcome confirms that the rule to decide the amount of text to be extracted from texts substantially depends for different writing styles.

\begin{table}[!t]
\centering
\subfloat[Performance of KEA]{
\begin{tabular}{ccccccc}
\hline 
\multirow{2}{*}{Classifier} & \multicolumn{2}{c}{FAO-780} & \multicolumn{2}{c}{CERN-290} & \multicolumn{2}{c}{NLM-500}\tabularnewline
 & F-Score & t-value & F-Score & t-value & F-Score & t-value\tabularnewline
\hline 
with Text Denoising & 23.03 & \multirow{2}{*}{5.07} & 14.73 & \multirow{2}{*}{3.42} & 14.60 & \multirow{2}{*}{4.14}\tabularnewline
Benchmark & 20.76 &  & 12.29 &  & 12.21 & \tabularnewline
\hline 
\end{tabular}
\label{KEA}}


\subfloat[Performance of KEA++]{
\begin{tabular}{ccccccc}
\hline 
\multirow{2}{*}{Classifier} & \multicolumn{2}{c}{FAO-780} & \multicolumn{2}{c}{CERN-290} & \multicolumn{2}{c}{NLM-500}\tabularnewline
 & F-Score & t-value & F-Score & t-value & F-Score & t-value\tabularnewline
\hline 
with Text Denoising & 27.98 & \multirow{2}{*}{3.78} & 23.28 & \multirow{2}{*}{2.40} & 20.15 & \multirow{2}{*}{6.38}\tabularnewline
Benchmark & 25.19 &  & 21.04 &  & 17.91 & \tabularnewline
\hline 
\end{tabular}
\label{KEA++}
}


\subfloat[Performance of Maui]{
\begin{tabular}{ccccccc}
\hline 
\multirow{2}{*}{Classifier} & \multicolumn{2}{c}{FAO-780} & \multicolumn{2}{c}{CERN-290} & \multicolumn{2}{c}{NLM-500}\tabularnewline
 & F-Score & t-value & F-Score & t-value & F-Score & t-value\tabularnewline
\hline 
with Text Denoising & 31.87 & \multirow{2}{*}{2.76} & 24.42 & \multirow{2}{*}{2.26} & 31.50 & \multirow{2}{*}{3.52}\tabularnewline
Benchmark & 31.86 &  & 24.92 &  & 31.13 & \tabularnewline
\hline 
\end{tabular}
\label{Maui}}

\caption{F-Scores of the keyphrase indexers with text denoising and its benchmark on three datasets}
\label{indexer}

\end{table}

\section{Text Denoising on Extracting Protein Relations bearing Sentences}
\label{ppi}
In an attempt to eliminate the denoising threshold which depends on writing style (Section \ref{keyphrase}), I decided to develop a machine learning version of text denoising. The classification task in hand was to annotate sentences of a set of texts with either \textit{positive} or \textit{negative} labels based on the presence of protein interactions. The feature set chosen is composed of $35$ features like various parameters of readability indexes, term frequency, inverse sentence frequency, biomedical named entity, verbs and acronyms, stopwords, semantic words, and sentence positions. After applying a series of well known classifiers like Bayesian classifiers, Random Forest, SVM, AdaBoost, and Bagging, the classifier that performed best was chosen which is Bagging stacked with Random Forest. The corpora used for this experiment are BioNLP, BioDRB and FetchProt that contain over $85,000$ sentences. Two automated tools called RelEx \cite{Relex:2007} and WRelEx \cite{wrelex:2012} are used to assign binary labels to each sentence of these corpora depending on the presence of any protein relations. Having realized after this assignment that the classes are negatively skewed (almost doubled the positive labels), synthetic positive samples are produced using SMOTE \cite{smote:2002} where the minority class is over-sampled
by taking each minority class sample and introducing synthetic examples along the line
segments joining any/all of the \textit{k}, which is five in our setup, minority class nearest neighbors. From initial results, I found that many features were highly correlated with each other but had low correlation with the class. Therefore, I used a \textit{wrapper} method to select a set of bias-free features. However, I observed that this set of features varies for different corpora. Table \ref{ppi} shows the precision, recall and F-Score of text denoising in a 10-fold cross validation setup, considering the highly agreed upon annotation of RelEx and WRelEx as the gold standard. It can be noted that the outcome of this experiment without using SMOTE was not satisfactory as the F-scores were under $80$\%.

\begin{table}[t]
\centering
\begin{tabular}{cccc}
\hline 
Dataset & Precision & Recall & F-Score\tabularnewline
\hline
BioNLP & 82.5 & 87.8 & 85.1\tabularnewline
BioDRB & 84.7 & 91.1 & 87.8\tabularnewline
FetchProt & 90.8 & 89.2 & 90\tabularnewline
\hline 
\end{tabular}
\caption{Performance of text denoising on extracting protein relations bearing sentences against the gold standard}
\label{ppi}
\end{table}

\section{Conclusion and Future Work}
\label{conclusion}
The proposed text denoising method performed much the same on several tasks and kinds of texts: the reduction of texts according to the readability improved relation mining, keyphrase indexing and extracting sentences that describe protein relations. This result strongly suggests that sentences that are difficult to read are more information-rich than the rest. The effect of text denoising is yet to be examined for text categorization and summarization. I am currently investigating the effect of readability on e-mail spam detection. The results so far are interesting as I am labelling spam and ham based on the readability of e-mail text content only (i.e., without looking at the mail header). Also, I intend to train benchmark summarizers with denoised texts and see how they perform against gold standard summaries.

\bibliographystyle{abbrv}
\bibliography{rushdibibliography}
\end{document}